\documentclass[aps,prl,reprint,groupedaddress,floatfix,showpacs]{revtex4-1}
\usepackage{graphicx, amsmath}

\begin{document}

\title{Ultrafast exciton formation at the ZnO(10${\overline{\textbf{1}}}$0) surface}
\mbox{\centering with kind permission from the American Physical Society, published as \href{http://journals.aps.org/prl/abstract/10.1103/PhysRevLett.113.057602}{Phys. Rev. Lett. 113, 057602 (2014)}}
\author{J.-C.~Deinert}
\email[]{deinert@fhi-berlin.mpg.de}
\author{D.~Wegkamp}
\author{M.~Meyer}
\author{C.~Richter}
\author{M.~Wolf}
\author{J.~St\"{a}hler}

\affiliation{Fritz-Haber-Institut der Max-Planck-Gesellschaft, Abteilung Physikalische Chemie, Faradayweg 4-6, 14195 Berlin}

\date{\today}

\begin{abstract}
We study the ultrafast quasiparticle dynamics in and below the ZnO 
conduction band using femtosecond time-resolved two-photon photoelectron spectroscopy. Above band gap excitation causes hot electron relaxation by electron-phonon scattering down to the 
Fermi level $E_\text{F}$ followed by ultrafast (200~fs) formation of a surface exciton (SX). Transient 
screening of the Coulomb interaction reduces the SX formation probability at high excitation densities near the Mott limit. Located just below the surface, 
the SX are stable with regard to hydrogen-induced work function modifications and thus the ideal prerequisite for resonant energy transfer applications.

\end{abstract}


\maketitle

The technological importance of zinc oxide (ZnO) originates from its large 
direct band gap ($\sim3.4~\text{eV}$) and high bulk exciton binding energy 
(60~meV) \cite{Woll2007}, which make it a promising candidate for 
optoelectronic applications, using ZnO nanoparticles \cite{Richters2008,Slootsky2014,KlingshirnPSS},  
epilayers \cite{KlingshirnPSS}, and hybrid organic-inorganic systems \cite{Della2011,Xu2013,Slootsky2014}. 
In this context, the optical and electronic properties of ZnO \emph{surfaces} 
naturally play a significant role with regard to electronic 
and excitonic coupling with other materials, e.g., organic dye molecules. As dipole-dipole coupling is highly distance dependent \cite{Forster1948}, the rates for F\"{o}rster resonance energy transfer (FRET) between different materials depend strongly on the presence of exciton dead layers at the surface \cite{Egel1996,Fono2004b, Kili2011} or, on the contrary, the existence of surface excitons. The latter are a dominant species in photoluminescence (PL) studies of different types of ZnO nano\-structures due to the comparably large surface-to-bulk ratio \cite{Fono2004,Richters2008,Kuehn2013,Slootsky2014}. 
Despite the great scientific attention, detailed understanding of this technologically highly important species is still missing, as the systematic modification and characterization of nanoparticle surfaces is very challenging.  Because of the lack of surface sensitivity of optical probes and PL, SX could only once be identified at a ZnO \emph{single crystal} surface  \cite{Trav1990}. However, as the sample had been exposed to air, the surface condition was rather undefined. Neither observation nor systematic characterization of SX under ultrahigh vacuum (UHV) conditions, which provide reproducible surfaces at the nanoscale level \cite{Woll2007, Diebold}, is known to date, and thus the origin and character of this very relevant species remains vague.

In this Letter, we use femtosecond (fs) time-resolved two-photon photoelectron (2PPE) spectroscopy to 
investigate the ultrafast carrier and exciton dynamics at the single crystal
ZnO($10\overline{1}0$) surface under well-defined UHV conditions. Due to the intrinsic surface sensitivity of 
this technique we are able to observe the ultrafast relaxation of hot 
electrons in the conduction band and the subsequent formation an 
excitonic state near the surface on sub-picosecond timescales. High excitation densities near the Mott limit enhance the screening of 
the electron-hole (\emph{e-h}) Coulomb interaction and thus reduce the formation probability of this state. Remarkably, even strong modification of the (static) surface charge 
density by hydrogen termination does \emph{not} change the population 
dynamics of the SX, strongly suggesting that the exciton is localized in the \emph{sub}surface 
region. This adjacence of the exciton to the surface  (1--2~nm) combined with its stability with respect to surface modification demonstrates its direct relevance for applications of ZnO involving resonant energy transfer.

\begin{figure}[]
\includegraphics{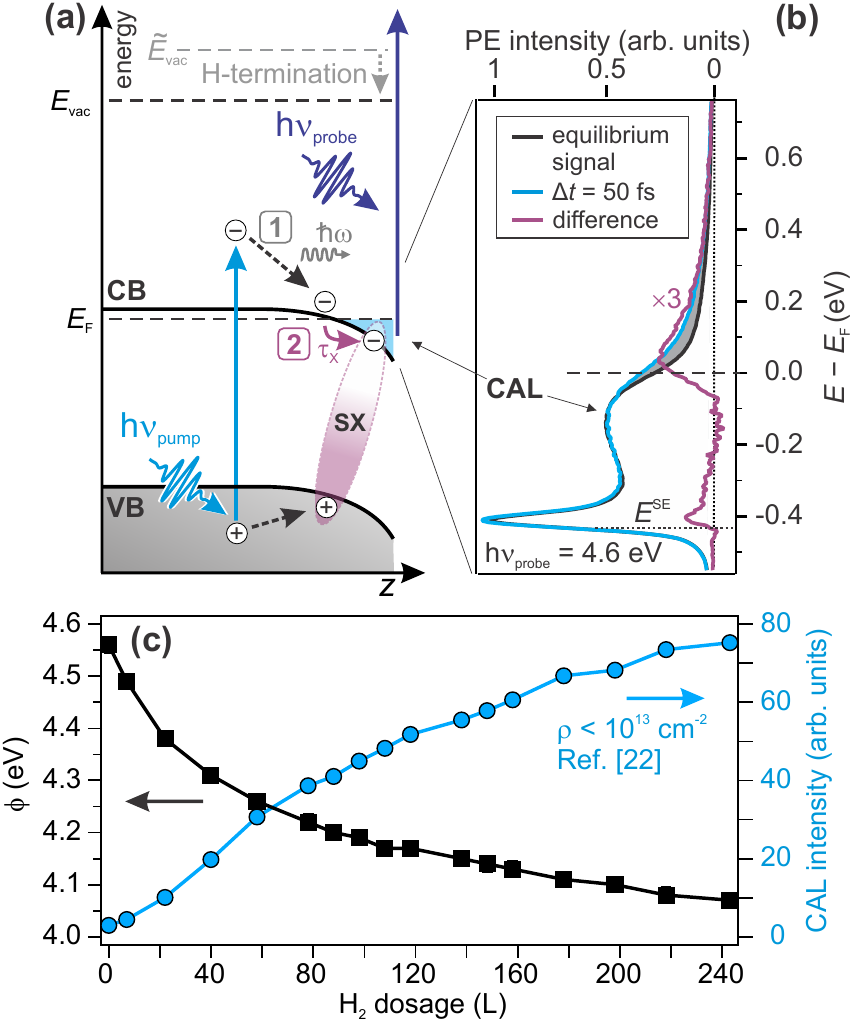}
\caption{\label{fig1}(color online). (a) 2PPE principle and relevant processes. Photoexcited ($h\nu_\text{pump}$) electrons relax by electron-phonon scattering (1) and exciton formation (2), which is probed 
by a time-delayed laser pulse ($h\nu_\text{probe}$) in photoemission. (b) Exemplary photoelectron spectrum of the hydrogen-terminated ZnO surface (black). The low-energy cut-off 
$E^\text{SE}=\Phi-h\nu_\text{probe}$ is a measure of the work function; the intensity 
below $E_\text{F}$ results from the CAL at the 
surface. Above band gap excitation leads to additional 
2PPE intensity around $E_\text{F}$ (blue). Subtraction of the equilibrium 
background signal yields the pump-induced spectrum (purple). (c) Impact of hydrogen termination on sample work function and 
CAL intensity.}
\end{figure}

Figure~\ref{fig1}(a) depicts the principle of time-resolved 2PPE spectroscopy. 
A first laser pulse initiates nonequilibrium dynamics by excitation of electrons from the valence band (VB) into normally unoccupied states in the conduction band (CB). A second laser pulse subsequently photoemits the excited electron population. The temporal evolution is monitored by 
variation of the time delay between pump and probe pulse \footnote{Laser pulses were 
provided by a regeneratively amplified (RegA), tuneable fs-laser (200~kHz). The  frequency-doubled output of an optical parametric amplifier 
provided $h\nu_\text{pump}=3.76\text{--}4.2~\text{eV}$ and the third (4.65~eV) 
and fourth harmonic (6.2~eV) of the 800~nm RegA output were used for 
photoemission}. The photoelectrons were detected using a hemispherical 
electron energy analyzer (PHOIBOS 100) held at a fixed bias voltage (1.5~V) with 
respect to the sample. Photoelectron (PE) spectra are referenced to
the, \textit{in situ} measured, tantalum sample holder Fermi energy $E_\text{F}$, which was in electrical contact 
with the sample surface.

We investigated three different hydrothermally grown ZnO single crystals (Mateck) and found no quantitative differences. Their nonpolar, mixed-terminated ($10\overline{1}0$) 
surfaces were prepared by $\text{Ar}^+$ sputtering and annealing cycles ($T_\text{max}
=850~\text{K}$) under UHV conditions ($p_\text{base}=1\times10^{-10}~\text{mbar}$).
The VB maximum remains unchanged at $E-E_\text{F}
=-3.18(6)~\text{eV}$ for different types of preparations using oxygen background pressures up to $10^{-6}\text{ mbar}$ (not shown). 
Neither charging, nor surface photovoltage effects have been observed \footnote{As discussed further below, the observed spectral features exhibit a neither fluence nor temperature dependence.}.
Unless otherwise stated, experiments were performed at $T=100~\text{K}$. Incident fluences between $10\text{ and }~35\mu\text{J/cm}^{2}$ translate to maximum excitation densities of $1.45-5.22\times10^{18}~\text{cm}^{-3}$, which is below or close to the Mott limit \footnote{Literature values for the Mott density vary between $1.5~\text{and}~6\times10^{18}~\text{cm}^{-3}$ \cite{Dijkhuis2011,Hendry2007,Bechstedt2011}. The calculated maximum excitation densities give an upper limit for the absorption directly at the surface.}. Hydrogen termination of the surface was achieved by background dosing of $
\text{H}_2$
\footnote{The glowing filament of an ion gauge was 
set $15~\text{cm}$ in line of sight with the sample surface leading to an increased dissociation of $\text{H}_2$.}.

Exposure of a semiconductor surface to an electron donor (like hydrogen)
can lead to downward surface band bending. In the case of the intrinsically 
\textit{n}-doped ZnO, this leads to the formation of a few $10~\text{\AA}$ thick 
charge accumulation layer (CAL) at the ($10\overline{1}0$) surface \cite{Eger1976,Lueth,Wang2005} with a 
maximum surface charge density of $10^{13}~\text{cm}^{-2}$ \cite{Ozawa2011}. We monitor the CAL formation by (1) the occurrence of photoelectron
intensity directly below $E_\text{F}$ as shown in Fig.~\ref{fig1}(b) and (2) by a reduction of the sample work function that is 
correlated with the surface band bending as sketched in Fig.~\ref{fig1}(a). 
Figure~\ref{fig1}(c) shows that increasing $\text{H}_2$ exposure leads, in agreement with literature \cite{Ozawa2011}, to an increasing CAL intensity (blue curve) \footnote{The sticking 
coefficient is unknown and dissociation of $\text{H}_2$ occurs at filaments in the chamber plays a role, which inhibits 
quantitative comparison of the exposures to Ref.~\cite{Ozawa2011}.}, which is accompanied by a work function decrease that saturates at $\Delta\Phi_\text{max}=-0.6~\text{eV}$. 
This change of the surface dipole is a result of the electron donor character of the 
hydrogen, leaving the positively charged ion at the surface. Directly after 
preparation of a pristine ZnO surface, a slight but continuous increase of 
CAL intensity going along with a work function reduction of several 10 meV is always observed 
due to the inevitable $\text{H}_2$ background even in a UHV environment 
($<0.1~\text{L/h}$). In order to stabilize the PE signal and to reduce the work function, we exposed the sample (unless otherwise stated) to $6~\text{L}$ 
of $\text{H}_2$ before experiments, which corresponds, based on Ref. \cite{Ozawa2011}, to a charge density in the CAL on the order of $10^{18}~\text{cm}^{-3}$ and a termination of surface oxygen of only a few percent.

\begin{figure*}
\includegraphics{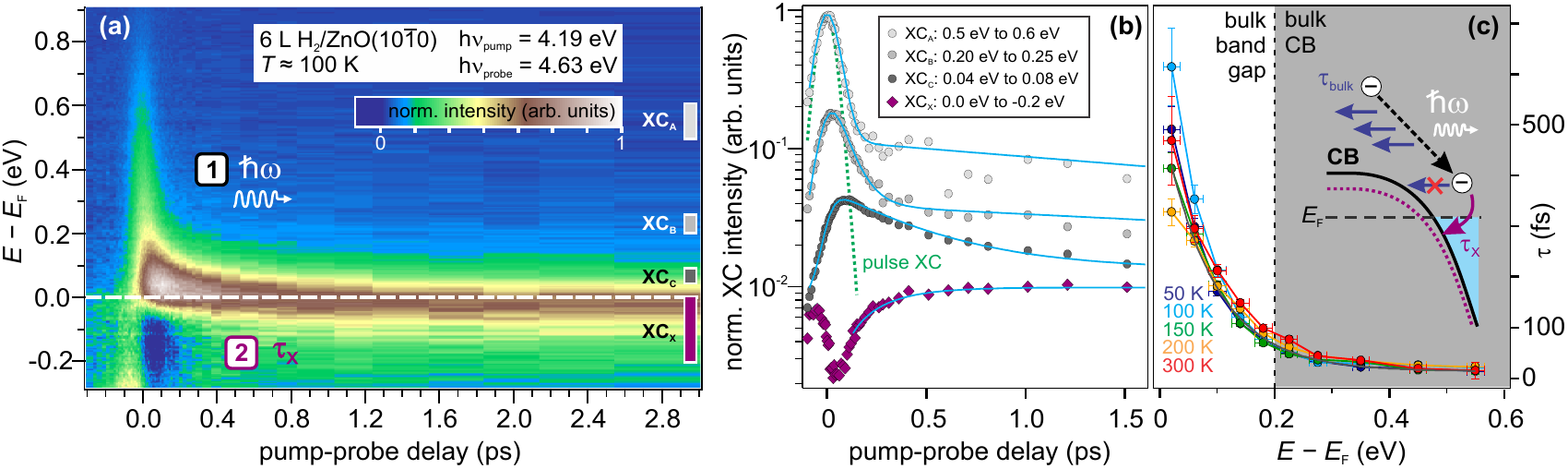}
\caption{\label{fig2}(color online). (a) Ultrafast electron dynamics at the ZnO surface below the Mott limit as 
probed by 2PPE spectroscopy. False colors represent  
photoelectrons created/depleted by $h\nu_\text{pump}$. Hot carriers in the CB relax on 
fs timescales by optical phonon emission. After a few 100~fs, 
additional electrons are verified below $E_\text{F}$, indicating that the 
surface exciton has formed. The signal at negative delays and energies 
results from photoelectrons with reversed excitation scheme (pumped by 4.6~eV and probed by 4.2~eV) and are a replica of the above dynamics shifted to 0.4~eV lower final state energies and evolving towards negative delays. (b) Pump and probe XC traces at indicated energies above (hot electrons) and below $E_\text{F}$ (exciton) and the corresponding double exponential fits to the data (solid lines). The instrument response function is represented by the dashed 
curve. (c) Temperature dependence of the hot electron relaxation time and 
scheme of the different relaxation channels.}
\end{figure*}

Figure~\ref{fig2}(a) shows the \emph{photoinduced} 2PPE intensity, i.e., after 
subtraction of the equilibrium background [cf. Fig.~\ref{fig1}(b)], in false colors as a function of pump-probe 
delay and energy with respect to $E_\text{F}$. Photoexcitation below the Mott limit (max. excitation density $1.45\times10^{18}~\text{cm}^{-3}$) launches fast dynamics at high energies  that significantly slow down for $E-E_\text{F} < 0.2~\text{eV}$.  Note that experiments at even lower excitation densities expose qualitatively
similar dynamics (not shown) \footnote{The observed dynamics thus cannot be attributed to collective phenomena of the excited electron-hole plasma like, e.g., Fermi edge singularities \cite{Kaindl1998,Perakis2000,Huber2001}.}. Exemplary cross correlation (XC) traces are 
depicted in Fig.~\ref{fig2}(b) (circles), which can be fitted (solid lines) using bi-exponential decays convolved with the laser pulses' cross correlation 
(dashed). The fast time constant of this empirical fit can be related to 
the hot electron relaxation time at the respective energy. We performed 
similar experiments for temperatures between 50 and 300~K, 
the results of which are presented in Fig.~\ref{fig2}(c). In agreement with 
previous 2PPE autocorrelation measurements \cite{Tisdale2008} and theoretical work \cite{Zhukov2010}, we find extremely fast electron 
relaxation for carriers above the bulk conduction band minimum (CBM) [$\tau=20\text{--}40~\text{fs}$,
gray shaded area in Fig.~\ref{fig2}(c)]. These fast relaxation times compared to hot electrons in metals \cite{Liso2004} can be explained by the reduced screening in the semiconductor. They result from 
scattering with optical phonons and are 
therefore independent of temperature in the investigated region ($k_
\text{B}T<26~\text{meV}$). Theory \cite{Zhukov2010} predicts a 
slowing down of the electron relaxation in the bulk when approaching the CBM [dashed line in Fig.~\ref{fig2}(c)], because fewer efficient scattering channels require more scattering events for the same energy loss. In the present experiment, at the 
ZnO($10\overline{1}0$) surface, such a ``phonon bottleneck" does not occur, as the downward surface band bending provides density of states (DOS) down to $E_
\text{F}$ [see inset in Fig. \ref{fig2}(c)].

The dynamics \emph{below} $E_\text{F}$ differ significantly from the ones above. 
Integration of the 2PPE intensity yields the diamond-shaped markers in Fig.~\ref{fig2}(b).
They display an initial drop of intensity, which is due to 
bleaching of the CAL by $h\nu_\text{pump}$, followed by a rise up to a constant positive
value. Notably, this delayed increase of pump-induced 2PPE intensity occurs 
\emph{below} $E_\text{F}$, i.e., below the energy which defines the 
highest occupied electronic state in equilibrium. Consequently, $h\nu_\text{pump}$ must create 
additional states below $E_\text{F}$ leading to an increased 2PPE signal. Such photoinduced creation of states could be caused by (i) small polaron 
formation, (ii) photoinduced changes of the surface electronic structure 
like surface photovoltage shifts, band gap renormalization, or heating/bleaching of the CAL, or (iii) exciton 
formation. 

\begin{figure}
\includegraphics{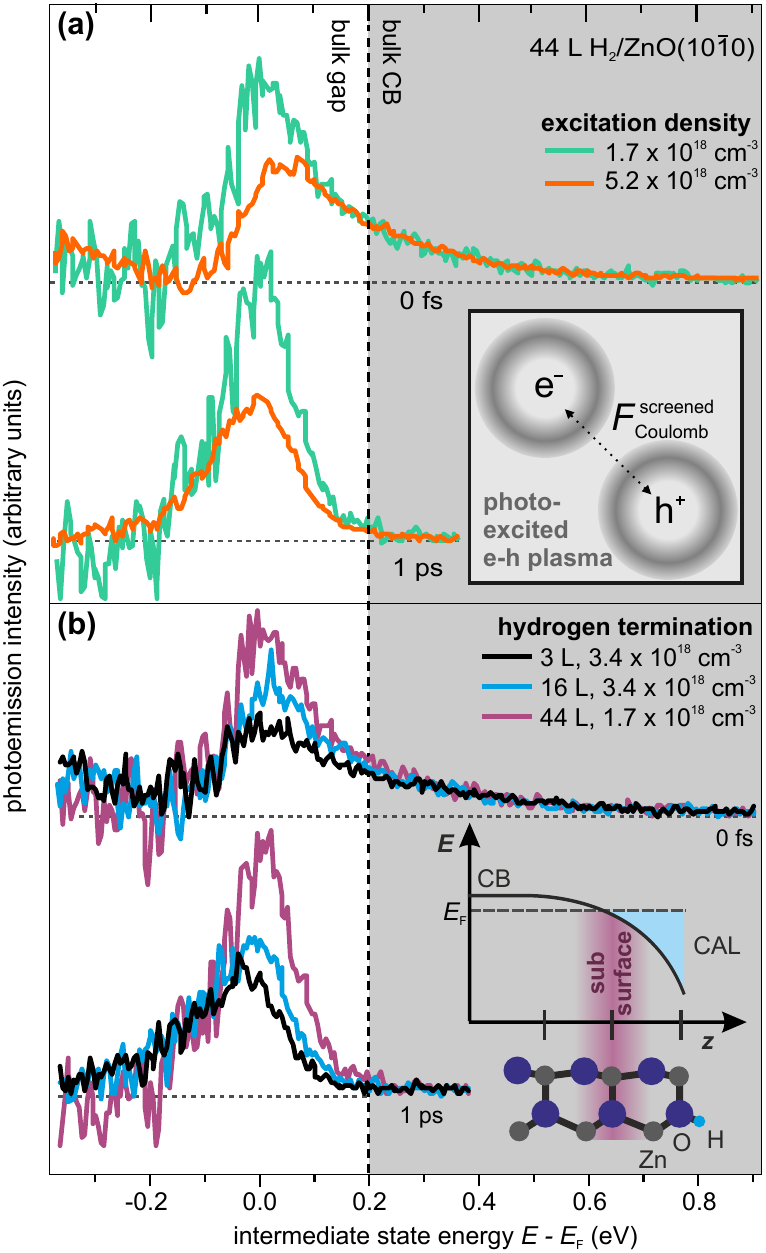}
\caption{\label{fig3}(color online). 2PPE spectra for (a) different excitation densities and (b) hydrogen terminations after background subtraction and normalized to the hot electron intensity at high energies ($0.4\text{--}0.6~\text{eV}$) 
in order to compare changes relative to the single particle excitation density. (a) Stronger excitation reduces the SX formation probability due to screening of the Coulomb interaction as illustrated in the insets. (b) The SX intensity does \emph{not} decrease with increasing $\text{H}_2$ termination 
showing localization in the subsurface region as illustrated by the inset.}
\end{figure}

It is well known that lattice distortions stabilize the ZnO exciton and lead to its large binding energy of 60~meV 
\footnote{The Coulomb attraction between electron 
and hole and its screening in the dielectric material determines the 
binding energy and localization of the exciton. In particular in 
semiconductors with ionic character like $\text{TiO}_2$ and ZnO, the 
coupling of carriers and excitons to the lattice, even on ultrafast 
timescales, cannot be neglected \cite{Chiodo2010,Zhukov2010,Bothschafter2013}. This leads to a polaronic character of the exciton (relaxation of ions localizes the \emph{e-h} pair), leading to a change of exciton binding energy due to coupling with phonons \cite{Hsu2004}.}. The formation of \emph{separate} polarons of electron and hole is thus unlikely. Also, no experimental
evidence of small polaron formation was found by recent time-resolved THz 
spectroscopy studies of ZnO \cite{Hendry2007}, excluding scenario (i). 
We tested the validity of (ii) and (iii) by a variation of the 
excitation density. While, in the case of (ii), photoinduced changes of the surface band bending (SBB) and 
occupational changes due to laser heating should become stronger with 
increasing pump fluence, (iii) the probability of exciton formation should be independent of excitation 
density below the Mott limit, i.e., the density at which the screening of the Coulomb interaction sets in [see inset in Fig.~\ref{fig3}(a)]. For stronger 
excitation the Coulomb interaction of the electron-hole pairs is reduced by the photoexcited \emph{e-h} plasma and thus the exciton 
formation probability diminishes.

Figure~\ref{fig3}(a) compares 2PPE spectra of two different excitation densities 
near the Mott limit of ZnO at different time delays.
While the overall dynamics do not change, the intensity of the spectral signature at $E_\text{F}$ is \emph{reduced} by the 
\emph{increase} of the excitation density. This observation unambiguously shows that (ii) photoinduced changes of SBB and CAL density cannot be the cause of the observed peak. On the contrary, the decreased intensity matches the expectations (iii) for an exciton whose formation probability is reduced by screening of the electron-hole interaction. As all of these processes must happen in the surface region \footnote{The SX feature is caused by the downward SBB, which extends only 1 nm into the ZnO crystal \cite{Ozawa2011}. Also, PE is only sensitive to the first few nm due to the finite mean free path of electrons in a solid.}, we conclude that the photoinduced intensity below $E_\text{F}$ results from near surface-bound excitons (SX) 
with a binding energy with respect to the bulk 
conduction band minimum of 250~meV. It should be noted that, in the SBB 
picture, the local binding energy relative to the local CBM is probably smaller [cf. inset of Fig.~\ref{fig2}(c), dotted curve]. The continuous evolution of the SX signature from the hot electron distribution is a direct consequence of the off-resonant excitation leading to carriers that relax to the band edges. The formation of SX occurs at least as fast as the observed rise time of PE intensity below $E_\text{F}$ of 200 fs, however, these excitons are in a highly excited "hot" state that relaxes to the ground state on longer timescales \footnote{The 1s ground state formation of the exciton usually occurs, depending on the material, on a ps to ns timescale. \cite{Chat2004,Hendry2007,Kaindl2009}}. Due to the limited energy resolution (45~meV) and because PES averages over the energy distribution in the SBB region, the spectral signatures of electron and exciton blend smoothly into each other \footnote{This is different to the SX formation at the Si(100) surface, where the exciton results from electrons and holes in the electronic surface states and not the CB and VB continuum \cite{Weinelt2004} and also not comparable to the transient excitons at the Ag(111) surface, which decay on fs timescales \cite{Petek2014}.}. As the electron of the SX energetically lies \emph{below} the Fermi level, nonradiative decay of this excitation is strongly suppressed: Dissociation of the exciton and separation of electron and hole in defect states within the band gap usually is a prominent nonradiative decay channel for excitons. It \emph{cannot} occur here, as most possible final states for the electron lie below $E_\text{F}$ and are therefore occupied. The decay of the SX population can thus only happen through electron-hole pair recombination going along with luminescence or Auger-type processes, which results in the high stability of this excited state \footnote{The SX population decay is slower than the 400~ps accessible in our experiment. Comparison with the results achieved by Tisdale \textit{et al.} \cite{Tisdale2008} using a 76~MHz oscillator suggests lifetimes exceeding 10~ns.}. 

In order to test the stability of the SX, we modified the surface 
electronic properties by variation of the hydrogen termination. As 
mentioned above, this leads to a reduction of the sample work function by 
hundreds of meV and concurrent formation of a CAL at the surface. Such 
strong modification of the surface electronic structure should severely 
affect the formation probability of the SX if it was localized in the 
vacuum in front of the surface (through the modified ionization potential 
leading to spectral shifts) or in the first double layer (through the 
screening of the Coulomb interaction by the charge density in the CAL leading to a reduction of the SX 
intensity with increasing termination). As shown in Fig.~\ref{fig3}(b), none of these effects occur when increasing the $ \text{H}_2$ exposure from 3 to 
44~L ($\Delta\Phi_\text{max}=-250~\text{meV}$). We thus conclude that the SX is localized in the subsurface region where additional static screening of the Coulomb interaction does not occur. On the contrary, increasing the hydrogen exposure by more than a factor of 10 leads to a comparably weak \emph{increase} of the SX intensity and a weak shift to higher energies. This observation could be explained by SX localized close to hydrogen binding sites where the downward SBB is strongest and a denser CAL with increasing termination that could give the SX Mahan exciton character (see e.g. \cite{Bechstedt2011}), i.e. the energetic stabilization of the CAL due to a localized hole in the VB.

In summary, we present a systematic investigation of the ultrafast electron 
and exciton dynamics at the ZnO($10\overline{1}0$) surface, showing the 
formation of a subsurface-bound exciton on fs timescales that 
exhibits a very large binding energy with respect to the bulk conduction 
band, resulting in a remarkable stability of this feature. The study thus offers a 
complete and novel picture of the quasiparticle relaxation at this surface. 
The existence of a subsurface exciton which is stable with regard to surface modifications is of high relevance for 
applications of ZnO, e.g., involving FRET.
The SX may dominate the dipole-dipole coupling across functional interfaces due to its adjacence to the surface, and it may even persist under non-UHV conditions. 

\begin{acknowledgments}
We are grateful for stimulating discussions with T. Kampfrath, P. 
Rinke and O. Hofmann at the FHI Berlin. This work was partially funded 
by the Deutsche Forschungsgemeinschaft through Sfb 951 and the European 
Union through Grant No. 280879-2 CRONOS CP-FP7.
\end{acknowledgments}

\bibliography{deinert_etal_ZnO}

\begin{thebibliography}{45}%
\makeatletter
\providecommand \@ifxundefined [1]{%
 \@ifx{#1\undefined}
}%
\providecommand \@ifnum [1]{%
 \ifnum #1\expandafter \@firstoftwo
 \else \expandafter \@secondoftwo
 \fi
}%
\providecommand \@ifx [1]{%
 \ifx #1\expandafter \@firstoftwo
 \else \expandafter \@secondoftwo
 \fi
}%
\providecommand \natexlab [1]{#1}%
\providecommand \enquote  [1]{``#1''}%
\providecommand \bibnamefont  [1]{#1}%
\providecommand \bibfnamefont [1]{#1}%
\providecommand \citenamefont [1]{#1}%
\providecommand \href@noop [0]{\@secondoftwo}%
\providecommand \href [0]{\begingroup \@sanitize@url \@href}%
\providecommand \@href[1]{\@@startlink{#1}\@@href}%
\providecommand \@@href[1]{\endgroup#1\@@endlink}%
\providecommand \@sanitize@url [0]{\catcode `\\12\catcode `\$12\catcode
  `\&12\catcode `\#12\catcode `\^12\catcode `\_12\catcode `\%12\relax}%
\providecommand \@@startlink[1]{}%
\providecommand \@@endlink[0]{}%
\providecommand \url  [0]{\begingroup\@sanitize@url \@url }%
\providecommand \@url [1]{\endgroup\@href {#1}{\urlprefix }}%
\providecommand \urlprefix  [0]{URL }%
\providecommand \Eprint [0]{\href }%
\providecommand \doibase [0]{http://dx.doi.org/}%
\providecommand \selectlanguage [0]{\@gobble}%
\providecommand \bibinfo  [0]{\@secondoftwo}%
\providecommand \bibfield  [0]{\@secondoftwo}%
\providecommand \translation [1]{[#1]}%
\providecommand \BibitemOpen [0]{}%
\providecommand \bibitemStop [0]{}%
\providecommand \bibitemNoStop [0]{.\EOS\space}%
\providecommand \EOS [0]{\spacefactor3000\relax}%
\providecommand \BibitemShut  [1]{\csname bibitem#1\endcsname}%
\let\auto@bib@innerbib\@empty
\bibitem [{\citenamefont {W{\"o}ll}(2007)}]{Woll2007}%
  \BibitemOpen
  \bibfield  {author} {\bibinfo {author} {\bibfnamefont {C.}~\bibnamefont
  {W{\"o}ll}},\ }\href {\doibase
  http://dx.doi.org/10.1016/j.progsurf.2006.12.002} {\bibfield  {journal}
  {\bibinfo  {journal} {Progress in Surface Science}\ }\textbf {\bibinfo
  {volume} {82}},\ \bibinfo {pages} {55} (\bibinfo {year} {2007})}\BibitemShut
  {NoStop}%
\bibitem [{\citenamefont {Richters}\ \emph {et~al.}(2008)\citenamefont
  {Richters}, \citenamefont {Voss}, \citenamefont {Kim}, \citenamefont
  {Scholz},\ and\ \citenamefont {Zacharias}}]{Richters2008}%
  \BibitemOpen
  \bibfield  {author} {\bibinfo {author} {\bibfnamefont {J.-P.}\ \bibnamefont
  {Richters}}, \bibinfo {author} {\bibfnamefont {T.}~\bibnamefont {Voss}},
  \bibinfo {author} {\bibfnamefont {D.~S.}\ \bibnamefont {Kim}}, \bibinfo
  {author} {\bibfnamefont {R.}~\bibnamefont {Scholz}}, \ and\ \bibinfo {author}
  {\bibfnamefont {M.}~\bibnamefont {Zacharias}},\ }\href {\doibase
  10.1088/0957-4484/19/30/305202} {\bibfield  {journal} {\bibinfo  {journal}
  {Nanotechnology}\ }\textbf {\bibinfo {volume} {19}},\ \bibinfo {pages}
  {305202} (\bibinfo {year} {2008})}\BibitemShut {NoStop}%
\bibitem [{\citenamefont {Slootsky}\ \emph {et~al.}(2014)\citenamefont
  {Slootsky}, \citenamefont {Liu}, \citenamefont {Menon},\ and\ \citenamefont
  {Forrest}}]{Slootsky2014}%
  \BibitemOpen
  \bibfield  {author} {\bibinfo {author} {\bibfnamefont {M.}~\bibnamefont
  {Slootsky}}, \bibinfo {author} {\bibfnamefont {X.}~\bibnamefont {Liu}},
  \bibinfo {author} {\bibfnamefont {V.~M.}\ \bibnamefont {Menon}}, \ and\
  \bibinfo {author} {\bibfnamefont {S.~R.}\ \bibnamefont {Forrest}},\ }\href
  {\doibase 10.1103/PhysRevLett.112.076401} {\bibfield  {journal} {\bibinfo
  {journal} {Physical Review Letters}\ }\textbf {\bibinfo {volume} {112}},\
  \bibinfo {pages} {076401} (\bibinfo {year} {2014})}\BibitemShut {NoStop}%
\bibitem [{\citenamefont {Klingshirn}\ \emph {et~al.}(2010)\citenamefont
  {Klingshirn}, \citenamefont {Fallert}, \citenamefont {Zhou}, \citenamefont
  {Sartor}, \citenamefont {Thiele}, \citenamefont {Maier-Flaig}, \citenamefont
  {Schneider},\ and\ \citenamefont {Kalt}}]{KlingshirnPSS}%
  \BibitemOpen
  \bibfield  {author} {\bibinfo {author} {\bibfnamefont {C.}~\bibnamefont
  {Klingshirn}}, \bibinfo {author} {\bibfnamefont {J.}~\bibnamefont {Fallert}},
  \bibinfo {author} {\bibfnamefont {H.}~\bibnamefont {Zhou}}, \bibinfo {author}
  {\bibfnamefont {J.}~\bibnamefont {Sartor}}, \bibinfo {author} {\bibfnamefont
  {C.}~\bibnamefont {Thiele}}, \bibinfo {author} {\bibfnamefont
  {F.}~\bibnamefont {Maier-Flaig}}, \bibinfo {author} {\bibfnamefont
  {D.}~\bibnamefont {Schneider}}, \ and\ \bibinfo {author} {\bibfnamefont
  {H.}~\bibnamefont {Kalt}},\ }\href {\doibase 10.1002/pssb.200983195}
  {\bibfield  {journal} {\bibinfo  {journal} {Physica Status Solidi (B)}\
  }\textbf {\bibinfo {volume} {247}},\ \bibinfo {pages} {1424} (\bibinfo {year}
  {2010})}\BibitemShut {NoStop}%
\bibitem [{\citenamefont {Della~Sala}\ \emph {et~al.}(2011)\citenamefont
  {Della~Sala}, \citenamefont {Blumstengel},\ and\ \citenamefont
  {Henneberger}}]{Della2011}%
  \BibitemOpen
  \bibfield  {author} {\bibinfo {author} {\bibfnamefont {F.}~\bibnamefont
  {Della~Sala}}, \bibinfo {author} {\bibfnamefont {S.}~\bibnamefont
  {Blumstengel}}, \ and\ \bibinfo {author} {\bibfnamefont {F.}~\bibnamefont
  {Henneberger}},\ }\href {\doibase 10.1103/PhysRevLett.107.146401} {\bibfield
  {journal} {\bibinfo  {journal} {Phys. Rev. Lett.}\ }\textbf {\bibinfo
  {volume} {107}},\ \bibinfo {pages} {146401} (\bibinfo {year}
  {2011})}\BibitemShut {NoStop}%
\bibitem [{\citenamefont {Xu}\ \emph {et~al.}(2013)\citenamefont {Xu},
  \citenamefont {Hofmann}, \citenamefont {Schlesinger}, \citenamefont
  {Winkler}, \citenamefont {Frisch}, \citenamefont {Niederhausen},
  \citenamefont {Vollmer}, \citenamefont {Blumstengel}, \citenamefont
  {Henneberger}, \citenamefont {Koch}, \citenamefont {Rinke},\ and\
  \citenamefont {Scheffler}}]{Xu2013}%
  \BibitemOpen
  \bibfield  {author} {\bibinfo {author} {\bibfnamefont {Y.}~\bibnamefont
  {Xu}}, \bibinfo {author} {\bibfnamefont {O.~T.}\ \bibnamefont {Hofmann}},
  \bibinfo {author} {\bibfnamefont {R.}~\bibnamefont {Schlesinger}}, \bibinfo
  {author} {\bibfnamefont {S.}~\bibnamefont {Winkler}}, \bibinfo {author}
  {\bibfnamefont {J.}~\bibnamefont {Frisch}}, \bibinfo {author} {\bibfnamefont
  {J.}~\bibnamefont {Niederhausen}}, \bibinfo {author} {\bibfnamefont
  {A.}~\bibnamefont {Vollmer}}, \bibinfo {author} {\bibfnamefont
  {S.}~\bibnamefont {Blumstengel}}, \bibinfo {author} {\bibfnamefont
  {F.}~\bibnamefont {Henneberger}}, \bibinfo {author} {\bibfnamefont
  {N.}~\bibnamefont {Koch}}, \bibinfo {author} {\bibfnamefont {P.}~\bibnamefont
  {Rinke}}, \ and\ \bibinfo {author} {\bibfnamefont {M.}~\bibnamefont
  {Scheffler}},\ }\href {\doibase 10.1103/PhysRevLett.111.226802} {\bibfield
  {journal} {\bibinfo  {journal} {Phys. Rev. Lett.}\ }\textbf {\bibinfo
  {volume} {111}},\ \bibinfo {pages} {226802} (\bibinfo {year}
  {2013})}\BibitemShut {NoStop}%
\bibitem [{\citenamefont {F\"{o}rster}(1948)}]{Forster1948}%
  \BibitemOpen
  \bibfield  {author} {\bibinfo {author} {\bibfnamefont {T.}~\bibnamefont
  {F\"{o}rster}},\ }\href {\doibase 10.1002/andp.19484370105} {\bibfield
  {journal} {\bibinfo  {journal} {Annalen der Physik}\ }\textbf {\bibinfo
  {volume} {437}},\ \bibinfo {pages} {55} (\bibinfo {year} {1948})}\BibitemShut
  {NoStop}%
\bibitem [{\citenamefont {Egelhaaf}\ and\ \citenamefont
  {Oelkrug}(1996)}]{Egel1996}%
  \BibitemOpen
  \bibfield  {author} {\bibinfo {author} {\bibfnamefont {H.-J.}\ \bibnamefont
  {Egelhaaf}}\ and\ \bibinfo {author} {\bibfnamefont {D.}~\bibnamefont
  {Oelkrug}},\ }\href {\doibase 10.1016/0022-0248(95)00634-6} {\bibfield
  {journal} {\bibinfo  {journal} {Journal of Crystal Growth}\ }\textbf
  {\bibinfo {volume} {161}},\ \bibinfo {pages} {190} (\bibinfo {year}
  {1996})}\BibitemShut {NoStop}%
\bibitem [{\citenamefont {Fonoberov}\ and\ \citenamefont
  {Balandin}(2004{\natexlab{a}})}]{Fono2004b}%
  \BibitemOpen
  \bibfield  {author} {\bibinfo {author} {\bibfnamefont {V.~A.}\ \bibnamefont
  {Fonoberov}}\ and\ \bibinfo {author} {\bibfnamefont {A.~A.}\ \bibnamefont
  {Balandin}},\ }\href {\doibase 10.1103/PhysRevB.70.195410} {\bibfield
  {journal} {\bibinfo  {journal} {Phys. Rev. B}\ }\textbf {\bibinfo {volume}
  {70}},\ \bibinfo {pages} {195410} (\bibinfo {year}
  {2004}{\natexlab{a}})}\BibitemShut {NoStop}%
\bibitem [{\citenamefont {Kiliani}\ \emph {et~al.}(2011)\citenamefont
  {Kiliani}, \citenamefont {Schneider}, \citenamefont {Litvinov}, \citenamefont
  {Gerthsen}, \citenamefont {Fonin}, \citenamefont {R\"{u}diger}, \citenamefont
  {Leitenstorfer},\ and\ \citenamefont {Bratschitsch}}]{Kili2011}%
  \BibitemOpen
  \bibfield  {author} {\bibinfo {author} {\bibfnamefont {G.}~\bibnamefont
  {Kiliani}}, \bibinfo {author} {\bibfnamefont {R.}~\bibnamefont {Schneider}},
  \bibinfo {author} {\bibfnamefont {D.}~\bibnamefont {Litvinov}}, \bibinfo
  {author} {\bibfnamefont {D.}~\bibnamefont {Gerthsen}}, \bibinfo {author}
  {\bibfnamefont {M.}~\bibnamefont {Fonin}}, \bibinfo {author} {\bibfnamefont
  {U.}~\bibnamefont {R\"{u}diger}}, \bibinfo {author} {\bibfnamefont
  {A.}~\bibnamefont {Leitenstorfer}}, \ and\ \bibinfo {author} {\bibfnamefont
  {R.}~\bibnamefont {Bratschitsch}},\ }\href {\doibase 10.1002/pssb.200983195}
  {\bibfield  {journal} {\bibinfo  {journal} {Optics Express}\ }\textbf
  {\bibinfo {volume} {19}},\ \bibinfo {pages} {1641} (\bibinfo {year}
  {2011})}\BibitemShut {NoStop}%
\bibitem [{\citenamefont {Fonoberov}\ and\ \citenamefont
  {Balandin}(2004{\natexlab{b}})}]{Fono2004}%
  \BibitemOpen
  \bibfield  {author} {\bibinfo {author} {\bibfnamefont {V.~A.}\ \bibnamefont
  {Fonoberov}}\ and\ \bibinfo {author} {\bibfnamefont {A.~A.}\ \bibnamefont
  {Balandin}},\ }\href {\doibase 10.1063/1.1835992} {\bibfield  {journal}
  {\bibinfo  {journal} {Applied Physics Letters}\ }\textbf {\bibinfo {volume}
  {85}},\ \bibinfo {pages} {5971} (\bibinfo {year}
  {2004}{\natexlab{b}})}\BibitemShut {NoStop}%
\bibitem [{\citenamefont {Kuehn}\ \emph {et~al.}(2013)\citenamefont {Kuehn},
  \citenamefont {Friede}, \citenamefont {Sadofev}, \citenamefont {Blumstengel},
  \citenamefont {Henneberger},\ and\ \citenamefont {Elsaesser}}]{Kuehn2013}%
  \BibitemOpen
  \bibfield  {author} {\bibinfo {author} {\bibfnamefont {S.}~\bibnamefont
  {Kuehn}}, \bibinfo {author} {\bibfnamefont {S.}~\bibnamefont {Friede}},
  \bibinfo {author} {\bibfnamefont {S.}~\bibnamefont {Sadofev}}, \bibinfo
  {author} {\bibfnamefont {S.}~\bibnamefont {Blumstengel}}, \bibinfo {author}
  {\bibfnamefont {F.}~\bibnamefont {Henneberger}}, \ and\ \bibinfo {author}
  {\bibfnamefont {T.}~\bibnamefont {Elsaesser}},\ }\href {\doibase
  http://dx.doi.org/10.1063/1.4829466} {\bibfield  {journal} {\bibinfo
  {journal} {Applied Physics Letters}\ }\textbf {\bibinfo {volume} {103}},\
  \bibinfo {eid} {191909} (\bibinfo {year} {2013})}\BibitemShut {NoStop}%
\bibitem [{\citenamefont {Travnikov}\ \emph {et~al.}(1990)\citenamefont
  {Travnikov}, \citenamefont {Freiberg},\ and\ \citenamefont
  {Savikhin}}]{Trav1990}%
  \BibitemOpen
  \bibfield  {author} {\bibinfo {author} {\bibfnamefont {V.}~\bibnamefont
  {Travnikov}}, \bibinfo {author} {\bibfnamefont {A.}~\bibnamefont {Freiberg}},
  \ and\ \bibinfo {author} {\bibfnamefont {S.}~\bibnamefont {Savikhin}},\
  }\href {\doibase 10.1016/0022-2313(90)90006-W} {\bibfield  {journal}
  {\bibinfo  {journal} {Journal of Luminescence}\ }\textbf {\bibinfo {volume}
  {47}},\ \bibinfo {pages} {107} (\bibinfo {year} {1990})}\BibitemShut
  {NoStop}%
\bibitem [{\citenamefont {Diebold}\ \emph {et~al.}(2004)\citenamefont
  {Diebold}, \citenamefont {Koplitz},\ and\ \citenamefont {Dulub}}]{Diebold}%
  \BibitemOpen
  \bibfield  {author} {\bibinfo {author} {\bibfnamefont {U.}~\bibnamefont
  {Diebold}}, \bibinfo {author} {\bibfnamefont {L.~V.}\ \bibnamefont
  {Koplitz}}, \ and\ \bibinfo {author} {\bibfnamefont {O.}~\bibnamefont
  {Dulub}},\ }\href {\doibase 10.1016/j.apsusc.2004.06.040} {\bibfield
  {journal} {\bibinfo  {journal} {Applied Surface Science}\ }\textbf {\bibinfo
  {volume} {237}},\ \bibinfo {pages} {336} (\bibinfo {year}
  {2004})}\BibitemShut {NoStop}%
\bibitem [{Note1()}]{Note1}%
  \BibitemOpen
  \bibinfo {note} {Laser pulses were provided by a regeneratively amplified
  (RegA), tuneable fs-laser (200~kHz). The frequency-doubled output of an
  optical parametric amplifier provided $h\nu _\protect \text
  {pump}=3.76\protect \text {--}4.2~\protect \text {eV}$ and the third
  (4.65~eV) and fourth harmonic (6.2~eV) of the 800~nm RegA output were used
  for photoemission}\BibitemShut {NoStop}%
\bibitem [{Note2()}]{Note2}%
  \BibitemOpen
  \bibinfo {note} {As discussed further below, the observed spectral features
  exhibit a neither fluence nor temperature dependence.}\BibitemShut {Stop}%
\bibitem [{Note3()}]{Note3}%
  \BibitemOpen
  \bibinfo {note} {Literature values for the Mott density vary between
  $1.5~\protect \text {and}~6\times 10^{18}~\protect \text {cm}^{-3}$ \cite
  {Dijkhuis2011,Hendry2007,Bechstedt2011}. The calculated maximum excitation
  densities give an upper limit for the absorption directly at the
  surface.}\BibitemShut {Stop}%
\bibitem [{Note4()}]{Note4}%
  \BibitemOpen
  \bibinfo {note} {The glowing filament of an ion gauge was set $15~\protect
  \text {cm}$ in line of sight with the sample surface leading to an increased
  dissociation of $\protect \text {H}_2$.}\BibitemShut {Stop}%
\bibitem [{\citenamefont {Eger}\ \emph {et~al.}(1976)\citenamefont {Eger},
  \citenamefont {Many},\ and\ \citenamefont {Goldstein}}]{Eger1976}%
  \BibitemOpen
  \bibfield  {author} {\bibinfo {author} {\bibfnamefont {D.}~\bibnamefont
  {Eger}}, \bibinfo {author} {\bibfnamefont {A.}~\bibnamefont {Many}}, \ and\
  \bibinfo {author} {\bibfnamefont {Y.}~\bibnamefont {Goldstein}},\ }\href
  {\doibase 10.1016/0039-6028(76)90107-2} {\bibfield  {journal} {\bibinfo
  {journal} {Surface Science}\ }\textbf {\bibinfo {volume} {58}},\ \bibinfo
  {pages} {18} (\bibinfo {year} {1976})}\BibitemShut {NoStop}%
\bibitem [{\citenamefont {L\"{u}th}(2010)}]{Lueth}%
  \BibitemOpen
  \bibfield  {author} {\bibinfo {author} {\bibfnamefont {H.}~\bibnamefont
  {L\"{u}th}},\ }\href {\doibase 10.1007/978-3-642-13592-7} {\emph {\bibinfo
  {title} {Solid Surfaces, Interfaces and Thin Films}}},\ edited by\ \bibinfo
  {editor} {\bibfnamefont {W.~T.}\ \bibnamefont {Rhodes}}, \bibinfo {editor}
  {\bibfnamefont {H.~E.}\ \bibnamefont {Stanley}}, \ and\ \bibinfo {editor}
  {\bibfnamefont {R.}~\bibnamefont {Needs}}\ (\bibinfo  {publisher} {Springer
  Berlin Heidelberg},\ \bibinfo {year} {2010})\BibitemShut {NoStop}%
\bibitem [{\citenamefont {Wang}\ \emph {et~al.}(2005)\citenamefont {Wang},
  \citenamefont {Meyer}, \citenamefont {Yin}, \citenamefont {Kunat},
  \citenamefont {Langenberg}, \citenamefont {Traeger}, \citenamefont
  {Birkner},\ and\ \citenamefont {W\"oll}}]{Wang2005}%
  \BibitemOpen
  \bibfield  {author} {\bibinfo {author} {\bibfnamefont {Y.}~\bibnamefont
  {Wang}}, \bibinfo {author} {\bibfnamefont {B.}~\bibnamefont {Meyer}},
  \bibinfo {author} {\bibfnamefont {X.}~\bibnamefont {Yin}}, \bibinfo {author}
  {\bibfnamefont {M.}~\bibnamefont {Kunat}}, \bibinfo {author} {\bibfnamefont
  {D.}~\bibnamefont {Langenberg}}, \bibinfo {author} {\bibfnamefont
  {F.}~\bibnamefont {Traeger}}, \bibinfo {author} {\bibfnamefont
  {A.}~\bibnamefont {Birkner}}, \ and\ \bibinfo {author} {\bibfnamefont
  {C.}~\bibnamefont {W\"oll}},\ }\href {\doibase 10.1103/PhysRevLett.95.266104}
  {\bibfield  {journal} {\bibinfo  {journal} {Phys. Rev. Lett.}\ }\textbf
  {\bibinfo {volume} {95}},\ \bibinfo {pages} {266104} (\bibinfo {year}
  {2005})}\BibitemShut {NoStop}%
\bibitem [{\citenamefont {Ozawa}\ and\ \citenamefont {Mase}(2011)}]{Ozawa2011}%
  \BibitemOpen
  \bibfield  {author} {\bibinfo {author} {\bibfnamefont {K.}~\bibnamefont
  {Ozawa}}\ and\ \bibinfo {author} {\bibfnamefont {K.}~\bibnamefont {Mase}},\
  }\href {\doibase 10.1103/PhysRevB.83.125406} {\bibfield  {journal} {\bibinfo
  {journal} {Phys. Rev. B}\ }\textbf {\bibinfo {volume} {83}},\ \bibinfo
  {pages} {125406} (\bibinfo {year} {2011})}\BibitemShut {NoStop}%
\bibitem [{Note5()}]{Note5}%
  \BibitemOpen
  \bibinfo {note} {The sticking coefficient is unknown and dissociation of
  $\protect \text {H}_2$ occurs at filaments in the chamber plays a role, which
  inhibits quantitative comparison of the exposures to Ref.~\cite
  {Ozawa2011}.}\BibitemShut {Stop}%
\bibitem [{Note6()}]{Note6}%
  \BibitemOpen
  \bibinfo {note} {The observed dynamics thus cannot be attributed to
  collective phenomena of the excited electron-hole plasma like, e.g., Fermi
  edge singularities \cite {Kaindl1998,Perakis2000,Huber2001}.}\BibitemShut
  {Stop}%
\bibitem [{\citenamefont {Tisdale}\ \emph {et~al.}(2008)\citenamefont
  {Tisdale}, \citenamefont {Muntwiler}, \citenamefont {Norris}, \citenamefont
  {Aydil},\ and\ \citenamefont {Zhu}}]{Tisdale2008}%
  \BibitemOpen
  \bibfield  {author} {\bibinfo {author} {\bibfnamefont {W.~A.}\ \bibnamefont
  {Tisdale}}, \bibinfo {author} {\bibfnamefont {M.}~\bibnamefont {Muntwiler}},
  \bibinfo {author} {\bibfnamefont {D.~J.}\ \bibnamefont {Norris}}, \bibinfo
  {author} {\bibfnamefont {E.~S.}\ \bibnamefont {Aydil}}, \ and\ \bibinfo
  {author} {\bibfnamefont {X.-Y.}\ \bibnamefont {Zhu}},\ }\href {\doibase
  10.1021/jp802455p} {\bibfield  {journal} {\bibinfo  {journal} {The Journal of
  Physical Chemistry C}\ }\textbf {\bibinfo {volume} {112}},\ \bibinfo {pages}
  {14682} (\bibinfo {year} {2008})}\BibitemShut {NoStop}%
\bibitem [{\citenamefont {Zhukov}\ \emph {et~al.}(2010)\citenamefont {Zhukov},
  \citenamefont {Echenique},\ and\ \citenamefont {Chulkov}}]{Zhukov2010}%
  \BibitemOpen
  \bibfield  {author} {\bibinfo {author} {\bibfnamefont {V.~P.}\ \bibnamefont
  {Zhukov}}, \bibinfo {author} {\bibfnamefont {P.~M.}\ \bibnamefont
  {Echenique}}, \ and\ \bibinfo {author} {\bibfnamefont {E.~V.}\ \bibnamefont
  {Chulkov}},\ }\href {\doibase 10.1103/PhysRevB.82.094302} {\bibfield
  {journal} {\bibinfo  {journal} {Phys. Rev. B}\ }\textbf {\bibinfo {volume}
  {82}},\ \bibinfo {pages} {094302} (\bibinfo {year} {2010})}\BibitemShut
  {NoStop}%
\bibitem [{\citenamefont {Lisowski}\ \emph {et~al.}(2004)\citenamefont
  {Lisowski}, \citenamefont {Loukakos}, \citenamefont {Bovensiepen},
  \citenamefont {St\"ahler}, \citenamefont {Gahl},\ and\ \citenamefont
  {Wolf}}]{Liso2004}%
  \BibitemOpen
  \bibfield  {author} {\bibinfo {author} {\bibfnamefont {M.}~\bibnamefont
  {Lisowski}}, \bibinfo {author} {\bibfnamefont {P.}~\bibnamefont {Loukakos}},
  \bibinfo {author} {\bibfnamefont {U.}~\bibnamefont {Bovensiepen}}, \bibinfo
  {author} {\bibfnamefont {J.}~\bibnamefont {St\"ahler}}, \bibinfo {author}
  {\bibfnamefont {C.}~\bibnamefont {Gahl}}, \ and\ \bibinfo {author}
  {\bibfnamefont {M.}~\bibnamefont {Wolf}},\ }\href {\doibase
  10.1007/s00339-003-2301-7} {\bibfield  {journal} {\bibinfo  {journal}
  {Applied Physics A}\ }\textbf {\bibinfo {volume} {78}},\ \bibinfo {pages}
  {165} (\bibinfo {year} {2004})}\BibitemShut {NoStop}%
\bibitem [{Note7()}]{Note7}%
  \BibitemOpen
  \bibinfo {note} {The Coulomb attraction between electron and hole and its
  screening in the dielectric material determines the binding energy and
  localization of the exciton. In particular in semiconductors with ionic
  character like $\protect \text {TiO}_2$ and ZnO, the coupling of carriers and
  excitons to the lattice, even on ultrafast timescales, cannot be neglected
  \cite {Chiodo2010,Zhukov2010,Bothschafter2013}. This leads to a polaronic
  character of the exciton (relaxation of ions localizes the \protect \emph
  {e-h} pair), leading to a change of exciton binding energy due to coupling
  with phonons \cite {Hsu2004}.}\BibitemShut {Stop}%
\bibitem [{\citenamefont {Hendry}\ \emph {et~al.}(2007)\citenamefont {Hendry},
  \citenamefont {Koeberg},\ and\ \citenamefont {Bonn}}]{Hendry2007}%
  \BibitemOpen
  \bibfield  {author} {\bibinfo {author} {\bibfnamefont {E.}~\bibnamefont
  {Hendry}}, \bibinfo {author} {\bibfnamefont {M.}~\bibnamefont {Koeberg}}, \
  and\ \bibinfo {author} {\bibfnamefont {M.}~\bibnamefont {Bonn}},\ }\href
  {\doibase 10.1103/PhysRevB.76.045214} {\bibfield  {journal} {\bibinfo
  {journal} {Phys. Rev. B}\ }\textbf {\bibinfo {volume} {76}},\ \bibinfo
  {pages} {045214} (\bibinfo {year} {2007})}\BibitemShut {NoStop}%
\bibitem [{Note8()}]{Note8}%
  \BibitemOpen
  \bibinfo {note} {The SX feature is caused by the downward SBB, which extends
  only 1 nm into the ZnO crystal \cite {Ozawa2011}. Also, PE is only sensitive
  to the first few nm due to the finite mean free path of electrons in a
  solid.}\BibitemShut {Stop}%
\bibitem [{Note9()}]{Note9}%
  \BibitemOpen
  \bibinfo {note} {The 1s ground state formation of the exciton usually occurs,
  depending on the material, on a ps to ns timescale. \cite
  {Chat2004,Hendry2007,Kaindl2009}}\BibitemShut {NoStop}%
\bibitem [{Note10()}]{Note10}%
  \BibitemOpen
  \bibinfo {note} {This is different to the SX formation at the Si(100)
  surface, where the exciton results from electrons and holes in the electronic
  surface states and not the CB and VB continuum \cite {Weinelt2004} and also
  not comparable to the transient excitons at the Ag(111) surface, which decay
  on fs timescales \cite {Petek2014}.}\BibitemShut {Stop}%
\bibitem [{Note11()}]{Note11}%
  \BibitemOpen
  \bibinfo {note} {The SX population decay is slower than the 400~ps accessible
  in our experiment. Comparison with the results achieved by Tisdale \protect
  \textit {et al.} \cite {Tisdale2008} using a 76~MHz oscillator suggests
  lifetimes exceeding 10~ns.}\BibitemShut {Stop}%
\bibitem [{\citenamefont {Schleife}\ \emph {et~al.}(2011)\citenamefont
  {Schleife}, \citenamefont {R{\"o}dl}, \citenamefont {Fuchs}, \citenamefont
  {Hannewald},\ and\ \citenamefont {Bechstedt}}]{Bechstedt2011}%
  \BibitemOpen
  \bibfield  {author} {\bibinfo {author} {\bibfnamefont {A.}~\bibnamefont
  {Schleife}}, \bibinfo {author} {\bibfnamefont {C.}~\bibnamefont {R{\"o}dl}},
  \bibinfo {author} {\bibfnamefont {F.}~\bibnamefont {Fuchs}}, \bibinfo
  {author} {\bibfnamefont {K.}~\bibnamefont {Hannewald}}, \ and\ \bibinfo
  {author} {\bibfnamefont {F.}~\bibnamefont {Bechstedt}},\ }\href@noop {}
  {\bibfield  {journal} {\bibinfo  {journal} {Phys. Rev. Lett.}\ }\textbf
  {\bibinfo {volume} {107}},\ \bibinfo {pages} {236405} (\bibinfo {year}
  {2011})}\BibitemShut {NoStop}%
\bibitem [{\citenamefont {Versteegh}\ \emph {et~al.}(2011)\citenamefont
  {Versteegh}, \citenamefont {Kuis}, \citenamefont {Stoof},\ and\ \citenamefont
  {Dijkhuis}}]{Dijkhuis2011}%
  \BibitemOpen
  \bibfield  {author} {\bibinfo {author} {\bibfnamefont {M.~A.~M.}\
  \bibnamefont {Versteegh}}, \bibinfo {author} {\bibfnamefont {T.}~\bibnamefont
  {Kuis}}, \bibinfo {author} {\bibfnamefont {H.~T.~C.}\ \bibnamefont {Stoof}},
  \ and\ \bibinfo {author} {\bibfnamefont {J.~I.}\ \bibnamefont {Dijkhuis}},\
  }\href {\doibase 10.1103/PhysRevB.84.035207} {\bibfield  {journal} {\bibinfo
  {journal} {Phys. Rev. B}\ }\textbf {\bibinfo {volume} {84}},\ \bibinfo
  {pages} {035207} (\bibinfo {year} {2011})}\BibitemShut {NoStop}%
\bibitem [{\citenamefont {Kaindl}\ \emph {et~al.}(1998)\citenamefont {Kaindl},
  \citenamefont {Lutgen}, \citenamefont {Woerner}, \citenamefont {Elsaesser},
  \citenamefont {Nottelmann}, \citenamefont {Axt}, \citenamefont {Kuhn},
  \citenamefont {Hase},\ and\ \citenamefont {K\"unzel}}]{Kaindl1998}%
  \BibitemOpen
  \bibfield  {author} {\bibinfo {author} {\bibfnamefont {R.~A.}\ \bibnamefont
  {Kaindl}}, \bibinfo {author} {\bibfnamefont {S.}~\bibnamefont {Lutgen}},
  \bibinfo {author} {\bibfnamefont {M.}~\bibnamefont {Woerner}}, \bibinfo
  {author} {\bibfnamefont {T.}~\bibnamefont {Elsaesser}}, \bibinfo {author}
  {\bibfnamefont {B.}~\bibnamefont {Nottelmann}}, \bibinfo {author}
  {\bibfnamefont {V.~M.}\ \bibnamefont {Axt}}, \bibinfo {author} {\bibfnamefont
  {T.}~\bibnamefont {Kuhn}}, \bibinfo {author} {\bibfnamefont {A.}~\bibnamefont
  {Hase}}, \ and\ \bibinfo {author} {\bibfnamefont {H.}~\bibnamefont
  {K\"unzel}},\ }\href {\doibase 10.1103/PhysRevLett.80.3575} {\bibfield
  {journal} {\bibinfo  {journal} {Phys. Rev. Lett.}\ }\textbf {\bibinfo
  {volume} {80}},\ \bibinfo {pages} {3575} (\bibinfo {year}
  {1998})}\BibitemShut {NoStop}%
\bibitem [{\citenamefont {Perakis}\ and\ \citenamefont
  {Shahbazyan}(2000)}]{Perakis2000}%
  \BibitemOpen
  \bibfield  {author} {\bibinfo {author} {\bibfnamefont {I.}~\bibnamefont
  {Perakis}}\ and\ \bibinfo {author} {\bibfnamefont {T.}~\bibnamefont
  {Shahbazyan}},\ }\href {\doibase
  http://dx.doi.org/10.1016/S0167-5729(00)00009-1} {\bibfield  {journal}
  {\bibinfo  {journal} {Surface Science Reports}\ }\textbf {\bibinfo {volume}
  {40}},\ \bibinfo {pages} {1 } (\bibinfo {year} {2000})}\BibitemShut {NoStop}%
\bibitem [{\citenamefont {Huber}\ \emph {et~al.}(2001)\citenamefont {Huber},
  \citenamefont {Tauser}, \citenamefont {Brodschelm}, \citenamefont {Bichler},
  \citenamefont {Abstreiter},\ and\ \citenamefont {Leitenstorfer}}]{Huber2001}%
  \BibitemOpen
  \bibfield  {author} {\bibinfo {author} {\bibfnamefont {R.}~\bibnamefont
  {Huber}}, \bibinfo {author} {\bibfnamefont {F.}~\bibnamefont {Tauser}},
  \bibinfo {author} {\bibfnamefont {A.}~\bibnamefont {Brodschelm}}, \bibinfo
  {author} {\bibfnamefont {M.}~\bibnamefont {Bichler}}, \bibinfo {author}
  {\bibfnamefont {G.}~\bibnamefont {Abstreiter}}, \ and\ \bibinfo {author}
  {\bibfnamefont {A.}~\bibnamefont {Leitenstorfer}},\ }\href {\doibase
  10.1038/35104522} {\bibfield  {journal} {\bibinfo  {journal} {Nature}\
  }\textbf {\bibinfo {volume} {414}},\ \bibinfo {pages} {286} (\bibinfo {year}
  {2001})}\BibitemShut {NoStop}%
\bibitem [{\citenamefont {Chiodo}\ \emph {et~al.}(2010)\citenamefont {Chiodo},
  \citenamefont {Garc{\`i}a-Lastra}, \citenamefont {Iacomino}, \citenamefont
  {Ossicini}, \citenamefont {Zhao}, \citenamefont {Petek},\ and\ \citenamefont
  {Rubio}}]{Chiodo2010}%
  \BibitemOpen
  \bibfield  {author} {\bibinfo {author} {\bibfnamefont {L.}~\bibnamefont
  {Chiodo}}, \bibinfo {author} {\bibfnamefont {J.~M.}\ \bibnamefont
  {Garc{\`i}a-Lastra}}, \bibinfo {author} {\bibfnamefont {A.}~\bibnamefont
  {Iacomino}}, \bibinfo {author} {\bibfnamefont {S.}~\bibnamefont {Ossicini}},
  \bibinfo {author} {\bibfnamefont {J.}~\bibnamefont {Zhao}}, \bibinfo {author}
  {\bibfnamefont {H.}~\bibnamefont {Petek}}, \ and\ \bibinfo {author}
  {\bibfnamefont {A.}~\bibnamefont {Rubio}},\ }\href {\doibase
  10.1103/PhysRevB.82.045207} {\bibfield  {journal} {\bibinfo  {journal} {Phys.
  Rev. B}\ }\textbf {\bibinfo {volume} {82}},\ \bibinfo {pages} {045207}
  (\bibinfo {year} {2010})}\BibitemShut {NoStop}%
\bibitem [{\citenamefont {Bothschafter}\ \emph {et~al.}(2013)\citenamefont
  {Bothschafter}, \citenamefont {Paarmann}, \citenamefont {Zijlstra},
  \citenamefont {Karpowicz}, \citenamefont {Garcia}, \citenamefont
  {Kienberger},\ and\ \citenamefont {Ernstorfer}}]{Bothschafter2013}%
  \BibitemOpen
  \bibfield  {author} {\bibinfo {author} {\bibfnamefont {E.}~\bibnamefont
  {Bothschafter}}, \bibinfo {author} {\bibfnamefont {A.}~\bibnamefont
  {Paarmann}}, \bibinfo {author} {\bibfnamefont {E.}~\bibnamefont {Zijlstra}},
  \bibinfo {author} {\bibfnamefont {N.}~\bibnamefont {Karpowicz}}, \bibinfo
  {author} {\bibfnamefont {M.}~\bibnamefont {Garcia}}, \bibinfo {author}
  {\bibfnamefont {R.}~\bibnamefont {Kienberger}}, \ and\ \bibinfo {author}
  {\bibfnamefont {R.}~\bibnamefont {Ernstorfer}},\ }\href {\doibase
  10.1103/PhysRevLett.110.067402} {\bibfield  {journal} {\bibinfo  {journal}
  {Physical Review Letters}\ }\textbf {\bibinfo {volume} {110}},\ \bibinfo
  {pages} {067402} (\bibinfo {year} {2013})}\BibitemShut {NoStop}%
\bibitem [{\citenamefont {Hsu}\ and\ \citenamefont {Hsieh}(2004)}]{Hsu2004}%
  \BibitemOpen
  \bibfield  {author} {\bibinfo {author} {\bibfnamefont {H.-C.}\ \bibnamefont
  {Hsu}}\ and\ \bibinfo {author} {\bibfnamefont {W.-F.}\ \bibnamefont
  {Hsieh}},\ }\href {\doibase http://dx.doi.org/10.1016/j.ssc.2004.05.043}
  {\bibfield  {journal} {\bibinfo  {journal} {Solid State Communications}\
  }\textbf {\bibinfo {volume} {131}},\ \bibinfo {pages} {371 } (\bibinfo {year}
  {2004})}\BibitemShut {NoStop}%
\bibitem [{\citenamefont {Chatterjee}\ \emph {et~al.}(2004)\citenamefont
  {Chatterjee}, \citenamefont {Ell}, \citenamefont {Mosor}, \citenamefont
  {Khitrova}, \citenamefont {Gibbs}, \citenamefont {Hoyer}, \citenamefont
  {Kira}, \citenamefont {Koch}, \citenamefont {Prineas},\ and\ \citenamefont
  {Stolz}}]{Chat2004}%
  \BibitemOpen
  \bibfield  {author} {\bibinfo {author} {\bibfnamefont {S.}~\bibnamefont
  {Chatterjee}}, \bibinfo {author} {\bibfnamefont {C.}~\bibnamefont {Ell}},
  \bibinfo {author} {\bibfnamefont {S.}~\bibnamefont {Mosor}}, \bibinfo
  {author} {\bibfnamefont {G.}~\bibnamefont {Khitrova}}, \bibinfo {author}
  {\bibfnamefont {H.~M.}\ \bibnamefont {Gibbs}}, \bibinfo {author}
  {\bibfnamefont {W.}~\bibnamefont {Hoyer}}, \bibinfo {author} {\bibfnamefont
  {M.}~\bibnamefont {Kira}}, \bibinfo {author} {\bibfnamefont {S.~W.}\
  \bibnamefont {Koch}}, \bibinfo {author} {\bibfnamefont {J.~P.}\ \bibnamefont
  {Prineas}}, \ and\ \bibinfo {author} {\bibfnamefont {H.}~\bibnamefont
  {Stolz}},\ }\href {\doibase 10.1103/PhysRevLett.92.067402} {\bibfield
  {journal} {\bibinfo  {journal} {Phys. Rev. Lett.}\ }\textbf {\bibinfo
  {volume} {92}},\ \bibinfo {pages} {067402} (\bibinfo {year}
  {2004})}\BibitemShut {NoStop}%
\bibitem [{\citenamefont {Kaindl}\ \emph {et~al.}(2009)\citenamefont {Kaindl},
  \citenamefont {H\"agele}, \citenamefont {Carnahan},\ and\ \citenamefont
  {Chemla}}]{Kaindl2009}%
  \BibitemOpen
  \bibfield  {author} {\bibinfo {author} {\bibfnamefont {R.~A.}\ \bibnamefont
  {Kaindl}}, \bibinfo {author} {\bibfnamefont {D.}~\bibnamefont {H\"agele}},
  \bibinfo {author} {\bibfnamefont {M.~A.}\ \bibnamefont {Carnahan}}, \ and\
  \bibinfo {author} {\bibfnamefont {D.~S.}\ \bibnamefont {Chemla}},\ }\href
  {\doibase 10.1103/PhysRevB.79.045320} {\bibfield  {journal} {\bibinfo
  {journal} {Phys. Rev. B}\ }\textbf {\bibinfo {volume} {79}},\ \bibinfo
  {pages} {045320} (\bibinfo {year} {2009})}\BibitemShut {NoStop}%
\bibitem [{\citenamefont {Weinelt}\ \emph {et~al.}(2004)\citenamefont
  {Weinelt}, \citenamefont {Kutschera}, \citenamefont {Fauster},\ and\
  \citenamefont {Rohlfing}}]{Weinelt2004}%
  \BibitemOpen
  \bibfield  {author} {\bibinfo {author} {\bibfnamefont {M.}~\bibnamefont
  {Weinelt}}, \bibinfo {author} {\bibfnamefont {M.}~\bibnamefont {Kutschera}},
  \bibinfo {author} {\bibfnamefont {T.}~\bibnamefont {Fauster}}, \ and\
  \bibinfo {author} {\bibfnamefont {M.}~\bibnamefont {Rohlfing}},\ }\href
  {\doibase 10.1103/PhysRevLett.92.126801} {\bibfield  {journal} {\bibinfo
  {journal} {Phys. Rev. Lett.}\ }\textbf {\bibinfo {volume} {92}},\ \bibinfo
  {pages} {126801} (\bibinfo {year} {2004})}\BibitemShut {NoStop}%
\bibitem [{\citenamefont {Cui}\ \emph {et~al.}(2014)\citenamefont {Cui},
  \citenamefont {Wang}, \citenamefont {Argondizzo}, \citenamefont
  {Garrett-Roe}, \citenamefont {Gumhalter},\ and\ \citenamefont
  {Petek}}]{Petek2014}%
  \BibitemOpen
  \bibfield  {author} {\bibinfo {author} {\bibfnamefont {X.}~\bibnamefont
  {Cui}}, \bibinfo {author} {\bibfnamefont {C.}~\bibnamefont {Wang}}, \bibinfo
  {author} {\bibfnamefont {A.}~\bibnamefont {Argondizzo}}, \bibinfo {author}
  {\bibfnamefont {S.}~\bibnamefont {Garrett-Roe}}, \bibinfo {author}
  {\bibfnamefont {B.}~\bibnamefont {Gumhalter}}, \ and\ \bibinfo {author}
  {\bibfnamefont {H.}~\bibnamefont {Petek}},\ }\href@noop {} {\bibfield
  {journal} {\bibinfo  {journal} {Nature Physics}\ }\textbf {\bibinfo {volume}
  {10}},\ \bibinfo {pages} {505} (\bibinfo {year} {2014})}\BibitemShut
  {NoStop}%
\end{thebibliography}%

\end{document}